\documentclass[seceq]{ptptex}

\usepackage{graphicx}

\title{
On the application of symmetry breaking and its restoration to treat pairing correlation in finite 
nuclei
}

\author{
Guillaume \textsc{Hupin}%
}

\inst{
$^*$ Grand Acc\'el\'erateur National d'Ions Lourds (GANIL), CEA/DSM-CNRS/IN2P3, Bvd Henri Becquerel, 14076 Caen, France\\
$^{\dag}$ Lawrence Livermore National Laboratory, P. O. Box 808, L-414, Livermore, California 94551, USA
}

\author{
Guillaume \textsc{Hupin}$^{*} \ ^{\dag}$, Denis \textsc{Lacroix}$^{*}$ %
}

\abst{
An alternative approach to symmetry restoration within Energy Density Functional, the Symmetry-Conserving 
EDF is discussed. In this approach, the energy is directly written in terms of the degrees of freedom  encoded in the one-, two-... body density matrices
of the state having good quantum numbers. The SC-EDF framework is illustrated within Projection After and Before variation 
applied to particle number restoration.  
}

\begin{document}

\maketitle

\section{Introduction}

The most popular method to describe pairing correlations in many-body systems is based on the
 Bardeen-Cooper-Schrieffer (BCS) or Hartree-Fock-Bogoliubov (HFB) approaches where one uses quasi-particle states 
 that explicitly break the U(1) symmetry associated with particle number conservation. This approach is 
extensively used in Energy Density Functional (EDF) theory. It is, however, known that the BCS or HFB
description of pairing 
 can be greatly improved, especially in the regime of weak coupling and/or small particle number, 
 if the particle number is properly restored \cite{Rin80}.  Such a restoration is usually made using configuration mixing 
 technique corresponding to a so-called Multi-Reference EDF (MR-EDF). 
 
BCS, HFB and symmetry restoration are well-defined techniques when using a Hamiltonian operator.
Additional care has to be taken when using a more general EDF eventually based on effective interactions.
The validity and range of applicability of configuration mixing within EDF methods 
 has been debated recently\cite{Ang01,Dob07}. In particular, (i) the 
 necessity to regularize the MR-EDF\cite{Lac09,Ben09,Dug09}, (ii) the density to be used in density dependent interaction \cite{Rob10}, and (iii) the need to clarify the notion of breaking/restoration of symmetries in density  functional theory \cite{Dug10} have been discussed.   
 
 Here,  the restoration of particle number within Energy Density Functional theory is re-analyzed. We show that the 
 MR-EDF can, under certain conditions, be interpreted as a functional theory of the one-body and two-body density matrices
 of the projected product state. Starting from this observation, we propose a new density functional approach, 
 which we call Symmetry-Conserving EDF (SC-EDF) where the breaking and restoration of symmetries are accounted for in a 
 single step \cite{Hup11}. This approach has the advantage to provide a suitable framework that solves the difficulties (i-iii) and to be 
 directly applicable to nuclei using most popular effective interaction. Applications to light and medium mass nuclei 
 of Projection After Variation (PAV) and Variation After Projection (VAP) will be presented.   
 
\section{The nuclear Energy Density Functional}

In Density Functional Theory (DFT) or Energy Density Functional (EDF) applied to nuclei, one generally starts from  a set of trial 
quasi-particle states, and writes the energy in terms of underlying densities. For instance, for the 
effective Skyrme or Gogny interactions, the EDF can be written as:
\begin{eqnarray}
{\cal E}[\Phi_0] =&& \mathcal{E} \left[ \rho , \kappa, \kappa^* \right] = \sum_{i} t_{ii} \rho_{ii}+ \frac{1}{2} \sum_{i,j} \overline{v}_{ijij}^{\rho \rho}  \rho_{ii}\rho_{jj}  + \frac{1}{4} 
 \sum_{i,j} \overline{v}_{i\bar\imath j\bar\jmath }^{\kappa \kappa} \kappa_{i \bar\imath }^* \kappa_{\bar\jmath j} \, ,
\label{eq:denssr}
\end{eqnarray}
where $\rho$ and $\kappa$ are the normal and anomalous density of the state $\Phi_0$ respectively, 
whereas $\overline{v}^{\rho \rho}$ and $\overline{v}^{\kappa \kappa}$
are the effective, eventually density dependent,  kernels in the mean-field and pairing channel,
respectively. 

At present, EDF methods are the only theory that provides a 
unified framework to describe nuclear structure, reaction and thermodynamics. One of the keys for this success 
is the use of trial states that do explicitly break some of the symmetries of the underlying many body Hamiltonian: translational invariance, 
parity, U(1) symmetry associated to particle number, ... Thanks to symmetry breaking, correlations like pairing can be included with a rather  
simple functional. 

The EDF based on a single trial state (Single-Reference [SR] approach) can provide a suitable approach for many aspects of nuclear physics.
In applications to nuclear spectroscopy however, it is 
often necessary to have states with good quantum numbers. 

The natural way to tackle this problem is to restore symmetries that have been broken at the SR level. Restoration 
of broken symmetries within nuclear EDF is strongly guided from the configuration mixing (Generator Coordinate Method) approach 
generally used in the Hamiltonian case (see discussion in \cite{Lac09}). This leads to the so-called Multi-Reference concept (MR-EDF). 
Unless a Hamiltonian is used as a starting point 
to design the EDF and no approximation is made on the exchange term, 
the justification of GCM like approaches is unclear. Recently, many practical and formal 
difficulties have pointed out some weakness of MR-EDF as they are currently formulated\cite{Ang01,Dob07,Lac09,Ben09,Dug09}.   
In particular, after restoration, the new functional may contains spurious components that do not properly behave under the symmetry group 
transformations. This has been clearly illustrated for the case of 
Particle Number Restoration (PNR) case\cite{Dob07,Ben09} and further discussed
in ref. \cite{Dug10} for the case of rotational invariance.
  
\section{The Symmetry-Conserving EDF concept}

We recently have shown that restoration of broken symmetries can eventually be made using a slightly different strategy and introducing
the concept of Symmetry-Conserving EDF.  This strategy is presented here.  Let use consider an EDF given by Eq. (\ref{eq:denssr}) and assume 
that a given symmetry ${\cal S}$ (spherical symmetry, particle number, ...) is broken. Introducing the transformation associated with this symmetry, 
denoted  by $R(\Omega)$. This operator can for instance represent a rotation in 3D space for the case of broken spherical symmetry or 
rotation in gauge space for the case of the $U(1)$ symmetry associated with particle number. Starting from $\Phi_0$, the energy of 
any rotated state, denoted by $| \Phi(\Omega) \rangle= R(\Omega) | \Phi_0 \rangle$ is generally degenerated with the original energy 
${\cal E}[\Phi (\Omega)] = {\cal E}[\Phi_0] $. To restore broken symmetries, a new trial state is generally introduced that sums up 
the different orientation, i.e.
\begin{eqnarray}
| \Psi_{\cal S} \rangle &=& \hat P_{\cal S} | \Phi_0 \rangle =\int_\Omega d\Omega G(\Omega) R(\Omega)| \Phi_0 \rangle = 
\int_\Omega d\Omega G(\Omega) | \Phi(\Omega) \rangle 
\end{eqnarray}  
where $G(\Omega)$ are weight factors that depend on the symmetry to be restored, and 
$\hat P_{\cal S}$ is the projector on good quantum numbers associated with the symmetry.  

The standard approach, guided by the GCM technique, to restore symmetries within EDF is to  directly write the energy as
\begin{eqnarray}
{\cal E}_{\cal S} [\Psi_{\cal S}] \equiv   \int_{\Omega}  d\Omega \, {\cal E} \left[ \rho^{0\Omega} , \kappa^{0\Omega}, 
{\kappa^{\Omega 0 }}^\star \right]
{\cal N}_{\cal S} ({0, \Omega}) \, .  \label{eq:ekernel}
\end{eqnarray}
where ${\cal N}_{\cal S} ({0, \Omega}) $ is a factors depending on $G(\Omega)$, while $\rho^{0\Omega}$, $\kappa^{0\Omega}$, ... are the transition 
densities associated with the couple of states $\{ \Phi_0 , \Phi(\Omega) \}$. 
Defining the energy as in (\ref{eq:ekernel}) for an EDF with different vertices $\overline{v}^{\rho \rho}$ and $\overline{v}^{\kappa \kappa}$,
that in addition might not be completely antisymmetric and contain elaborate density dependencies, will lead to 
spurious 
terms that might eventually spoil MR-EDF applications.
 
Recently, we proposed to write the energy directly in terms of the degrees of freedom associated with the projected 
state\cite{Hup11}, i.e.
\begin{eqnarray}
{\cal E}_{\cal S} [\Psi_{\cal S}] \equiv   {\cal E}_{\cal S} [\rho^{\cal S}, R^{\cal S}, ...] .  \label{eq:ekernel2}
\end{eqnarray} 
where $\rho^{\cal S}$, $R^{\cal S}$, ... denote, respectively, the one-, two-, ... body densities of a symmetry restored state. Such an approach called 
hereafter symmetry conserving is motivated by the following reasons:
\begin{itemize}
  \item When a Hamiltonian is used equation (\ref{eq:ekernel2}) and (\ref{eq:ekernel}) are strictly equivalent \cite{Hup11-2}.
  \item For functionals that can be regularized, it has been shown recently that, in the case of PNR, the regularized MR-EDF takes a form 
  very close to (\ref{eq:ekernel2}) \cite{Hup11}.    
  \item Since the building block of the functional theory are observables of the projected state, they automatically include 
  the proper properties under the transformation of the restored symmetry and so do the energy after projection.
  \item The SC-EDF can be regarded as an extension of the SR-EDF onto a different, eventually larger Hilbert space of
  trial states.  
\end{itemize}

Illustrations of the SC-EDF application are given below for the PNR case. 

\section{Applications}

\begin{figure}[htb]
\begin{center}
\includegraphics[width=6.5cm]{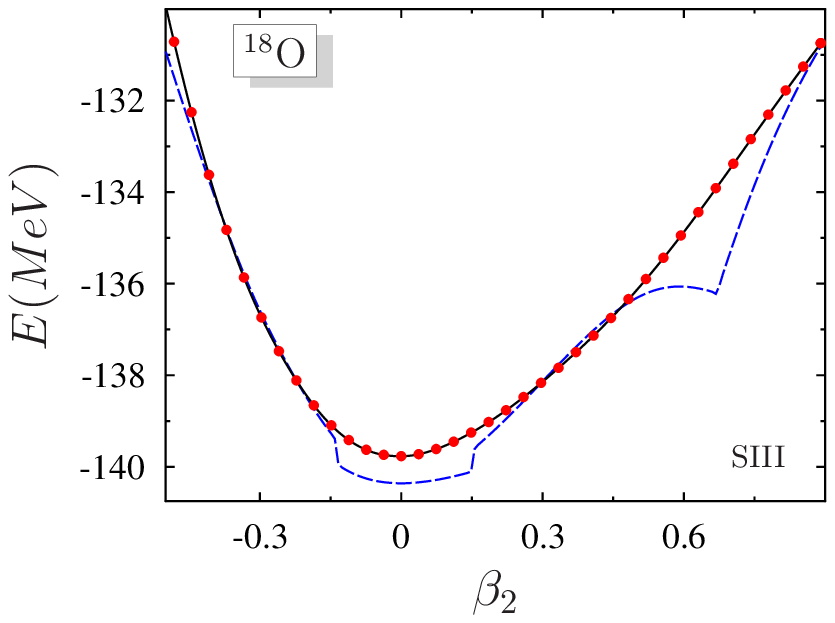}
\includegraphics[width=6.5cm]{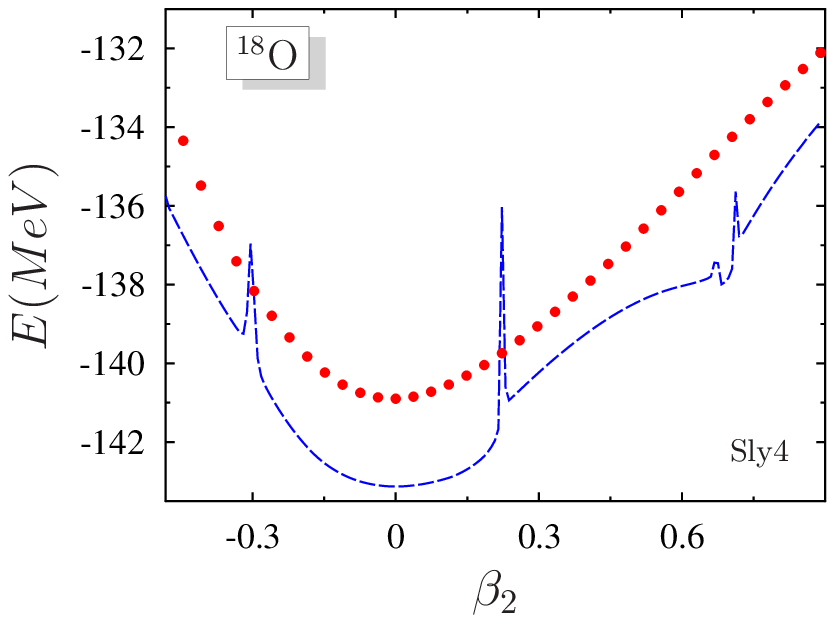}
\end{center}
\caption{(Color online) Left:
Particle-number restored deformation energy curve of $^{18}$O as a function of quadrupole 
deformation $\beta_2$ and calculated 
within standard MR-EDF technique using Projection 
After Variation (PAV) with the parameterization
SIII of the Skyrme EDF and a delta pairing interaction without (blue dashed curve) and with regularization (black solid curve).
The red filled circles correspond to the result 
obtained using directly Eq.~(\ref{eq:edftotprojcor}) (see text).
To compare with previous work \cite{Ben09}, the Coulomb exchange contribution has been subtracted from the energy.
Right: Same for the Sly$4$ parameterization in the mean-field channel. In that case, only two curves can be shown since the MR-EDF
cannot be regularized. To emphasize the steps and divergences, the MR-EDF calculations were performed with many more discretization 
points of the gauge-space integrals than usually done, cf ref. \cite{Ben09} for further details.   
}
\label{fig1:ykis}
\end{figure}

Starting from a general discussion on MR-EDF applied to particle number restoration, it has been 
shown that the SR-EDF can be conveniently extended to include the conservation of particle number. 
In that case, the energy reads
\begin{eqnarray}
 {\cal E}_{N} [\Psi_{N}] &=& \sum_{i} t_{ii} n^N_i 
+ \frac{1}{2} \sum_{i,j, j\neq \bar\imath } \overline{v}_{ijij}^{\rho \rho}   R^N_{ijij} 
+ \frac{1}{4} \sum_{i \neq j, j\neq \bar\imath } \overline{v}_{i\bar\imath j\bar\jmath }^{\kappa \kappa}  R^N_{j \bar\jmath  i\bar\imath  }  
\nonumber \\
&+& \frac{1}{2} \sum_{i} \overline{v}_{i\bar\imath  i \bar\imath }^{\rho \rho}  n^N_i n^N_i 
+\frac{1}{2} \sum_{i}\overline{v}_{i\bar\imath i\bar\imath }^{\kappa \kappa}  n^N_i (1 - n^N_i)  \,  , 
 \label{eq:edftotprojcor}
\end{eqnarray}  
where $n^N$ denotes the occupation numbers and $R^N$ the two-body density matrix of the state with particle number $N$, denoted by $| \Psi_N \rangle$. 

This SC-EDF has been used to perform Projection After Variation (PAV) \cite{Hup11} similarly to what is currently done in MR-EDF (see figure \ref{fig1:ykis}). 
When regularization of MR-EDF is possible, the energy obtained using Eq. (\ref{eq:edftotprojcor}) is strictly equivalent to the result of 
the regularized EDF (left of figure \ref{fig1:ykis}). Contrary to MR-EDF, the SC-EDF can also be applied to functional containing a non-integer 
powers of the density (right side of \ref{fig1:ykis}) leading to a smooth behavior of the potential energy curves.

\begin{figure}[htbp]
\begin{center}
\includegraphics[width=8.5cm]{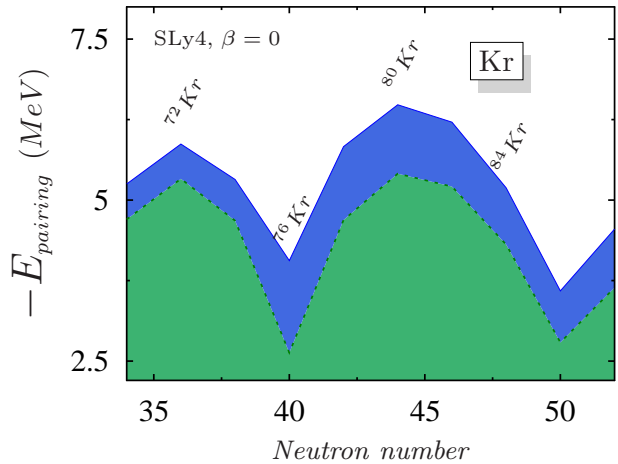}
\end{center}
\caption{(Color online) Illustration of the (proton+neutron) pairing energy obtained after PNR-VAP (solid curve). The dashed 
curve correspond to the original SR-EDF result. }
\label{fig3:ykis}
\end{figure}

More recently, the SC-EDF has been used to perform Variation After Projection (VAP). In that case, the energy is minimized directly
by making the variation of the projected state degrees of freedom. In figure \ref{fig3:ykis} an illustration of the energy gain in the pairing 
channel is shown for the krypton isotopic chain. One of the main advantages of VAP is to give also non-zero pairing energy 
for closed-shell nuclei. For instance, the $^{76}$Kr is at a neutron sub-shell closure $N=40$. For the pairing functional used here, 
HFB or BCS approaches lead to zero 
pairing energy in the neutron channel, whereas the VAP gives a pairing energy of 4.1 MeV. Note that this energy gain is partially compensated by 
the rearrangement of the mean-field, leading to a net gain of energy of 1 MeV.   

\section{Summary}

Motivated by recent difficulties observed in MR-EDF, we have recently introduced a new symmetry-conserved EDF framework. 
Application to particle number restoration are very encouraging. In particular, this approach can be applied to 
functionals with rather general dependence on the density, unlike standard regularized MR-EDF. In addition, its application 
to VAP, illustrates its feasibility. The main difficulty is now to provide functional 
forms that can be used for other symmetry restoration
like deformation.

\section*{Acknowledgments}
We would like to thank M. Bender for stimulating discussions and helpful advices.

%

\end{document}